\documentclass{ptephy}
\usepackage{graphicx}
\usepackage{lineno}
\usepackage{ulem}

\begin{document}

\title{Improved search for two-neutrino double electron capture on $^{124}$Xe and $^{126}$Xe
using particle identification in XMASS-I}

\newcommand{\ICRR}{1}
\newcommand{\IBS}{2}
\newcommand{\ISEE}{3}
\newcommand{\Tokushima}{4}
\newcommand{\IPMU}{5}
\newcommand{\KMI}{6}
\newcommand{\Kobe}{7}
\newcommand{\KRISS}{8}
\newcommand{\Miyagi}{9}
\newcommand{\Tokai}{10}
\newcommand{\YNU}{11}
\newcommand{\Tokushimanow}{\dagger}
\newcommand{\Tohokunow}{\ddagger}

\author{\name{{\bf XMASS Collaboration}}{\ast}\\
\name{K.~Abe}{\ICRR,\IPMU},
\name{K.~Hiraide}{\ICRR,\IPMU},
\name{K.~Ichimura}{\ICRR,\IPMU},
\name{Y.~Kishimoto}{\ICRR,\IPMU},
\name{K.~Kobayashi}{\ICRR,\IPMU},
\name{M.~Kobayashi}{\ICRR},
\name{S.~Moriyama}{\ICRR,\IPMU},
\name{M.~Nakahata}{\ICRR,\IPMU},
\name{T.~Norita}{\ICRR},
\name{H.~Ogawa}{\ICRR,\IPMU},
\name{K.~Sato}{\ICRR},
\name{H.~Sekiya}{\ICRR,\IPMU},
\name{O.~Takachio}{\ICRR},
\name{A.~Takeda}{\ICRR,\IPMU},
\name{S.~Tasaka}{\ICRR},
\name{M.~Yamashita}{\ICRR,\IPMU},
\name{B.~S.~Yang}{\ICRR,\IPMU},
\name{N.~Y.~Kim}{\IBS},
\name{Y.~D.~Kim}{\IBS},
\name{Y.~Itow}{\ISEE,\KMI},
\name{K.~Kanzawa}{\ISEE},
\name{R.~Kegasa}{\ISEE},
\name{K.~Masuda}{\ISEE},
\name{H.~Takiya}{\ISEE},
\name{K.~Fushimi}{\Tokushima,}\thanks{Now at Department of Physics, Tokushima University, 2-1 Minami Josanjimacho Tokushima city, Tokushima, 770-8506, Japan},
\name{G.~Kanzaki}{\Tokushima},
\name{K.~Martens}{\IPMU},
\name{Y.~Suzuki}{\IPMU},
\name{B.~D.~Xu}{\IPMU},
\name{R.~Fujita}{\Kobe},
\name{K.~Hosokawa}{\Kobe,}\thanks{Now at Research Center for Neutrino Science, Tohoku University, Sendai, Miyagi 980-8578, Japan},
\name{K.~Miuchi}{\Kobe},
\name{N.~Oka}{\Kobe},
\name{Y.~Takeuchi}{\Kobe,\IPMU},
\name{Y.~H.~Kim}{\KRISS,\IBS},
\name{K.~B.~Lee}{\KRISS},
\name{M.~K.~Lee}{\KRISS},
\name{Y.~Fukuda}{\Miyagi},
\name{M.~Miyasaka}{\Tokai},
\name{K.~Nishijima}{\Tokai},
\name{S.~Nakamura}{\YNU}
}

\address{
\affil{\ICRR}{Kamioka Observatory, Institute for Cosmic Ray Research, the University of Tokyo, Higashi-Mozumi, Kamioka, Hida, Gifu 506-1205, Japan}
\affil{\IBS}{Center of Underground Physics, Institute for Basic Science, 70 Yuseong-daero 1689-gil, Yuseong-gu, Daejeon 305-811, South Korea}
\affil{\ISEE}{Institute for Space-Earth Environmental Research, Nagoya University, Nagoya, Aichi 464-8601, Japan}
\affil{\Tokushima}{Institute of Socio-Arts and Sciences, The University of Tokushima, 1-1 Minamijosanjimacho Tokushima city, Tokushima, 770-8502, Japan}
\affil{\IPMU}{Kavli Institute for the Physics and Mathematics of the Universe (WPI), the University of Tokyo, Kashiwa, Chiba 277-8582, Japan}
\affil{\KMI}{Kobayashi-Maskawa Institute for the Origin of Particles and the Universe, Nagoya University, Furo-cho, Chikusa-ku, Nagoya, Aichi 464-8602, Japan}
\affil{\Kobe}{Department of Physics, Kobe University, Kobe, Hyogo 657-8501, Japan}
\affil{\KRISS}{Korea Research Institute of Standards and Science, Daejeon 305-340, South Korea}
\affil{\Miyagi}{Department of Physics, Miyagi University of Education, Sendai, Miyagi 980-0845, Japan}
\affil{\Tokai}{Department of Physics, Tokai University, Hiratsuka, Kanagawa 259-1292, Japan}
\affil{\YNU}{Department of Physics, Faculty of Engineering, Yokohama National University, Yokohama, Kanagawa 240-8501, Japan}
\email{xmass.publications7@km.icrr.u-tokyo.ac.jp}}


\begin{abstract}%
We conducted an improved search for the simultaneous capture of two $K$-shell electrons
on the $^{124}$Xe and $^{126}$Xe nuclei with emission of two neutrinos
using 800.0 days of data from the XMASS-I detector.
A novel method to discriminate $\gamma$-ray/$X$-ray or double electron capture signals
from $\beta$-ray background using scintillation time profiles was developed for this search.
No significant signal was found when fitting the observed
energy spectra with the expected signal and background.
Therefore, we set the most stringent lower limits on
the half-lives at $2.1 \times 10^{22}$ and $1.9 \times 10^{22}$ years 
for $^{124}$Xe and  $^{126}$Xe, respectively, with 90\% confidence level.
These limits improve upon previously reported values by a factor of 4.5.
\end{abstract}

\subjectindex{C04, C43, D29}

\maketitle

\section{Introduction}
Double electron capture (ECEC) is a rare nuclear decay process where a nucleus captures two orbital electrons
simultaneously. There might be two modes of the process:
\begin{eqnarray}
    (Z,A) + 2e^{-} &\to& (Z-2, A)          \ , \\
    (Z,A) + 2e^{-} &\to& (Z-2, A) + 2\nu_e \ ,
\end{eqnarray}
where $Z$ and $A$ are the atomic number and atomic mass number of the nucleus, respectively.

Detecting the neutrinoless mode of this process (0$\nu$ECEC) would provide evidence for
lepton number violation and the Majorana nature of the neutrino if observed.
To release the decay energy in 0$\nu$ECEC, there are two proposed mechanisms:
the radiative and the resonant mechanisms.
In the case of the radiative mechanism, the decay energy is carried away
by emitting, for example, an internal Bremsstrahlung photon~\cite{Winter:1955zz,Doi:1992dm}.
This process is, however, expected to have a much longer life-time
than neutrinoless double beta decay.
On the other hand, an enhancement of the capture rate by a factor as large as $10^{6}$
is possible if the initial and final (excited) masses of the nucleus
are degenerate~\cite{Vergados:1982wr,Bernabeu:1983yb,Sujkowski:2003mb,Frekers:2005ze,Krivoruchenko:2010ng,Kotila:2014zya}.
Therefore, experimental searches for 0$\nu$ECEC have been recently performed
for a variety of candidate nuclei~\cite{Barabash:2006qx,Barabash:2009ja,Belli:2013qja,Belli:2014map,Gavrilyuk:2013yqa,
Finch:2015hly,Jeskovsky:2015jia,Angloher:2016ktr,Agostini:2016rsa,Lehnert:2016gra,
Lehnert:2016tlk}.

Although two-neutrino double electron capture (2$\nu$ECEC) is allowed within
the Standard Model of particle physics,
only a few positive experimental results for 2$\nu$ECEC have been reported:
geochemical measurements of $^{130}$Ba~\cite{Meshik:2001ra,Pujol:2009}
and a direct measurement of $^{78}$Kr~\cite{Gavrilyuk:2013yqa,Ratkevich:2017kaz}
with half-lives of the order of $10^{21}$--$10^{22}$~years.
Despite the nuclear matrix element for the two-neutrino mode differs from that for
the neutrinoless mode, they are related to each other through the relevant parameters
in a chosen nuclear model~\cite{Bilenky:2014uka}.
For instance, the nucleus' axial current coupling constant $g_A$ and
the strength of the particle-particle interaction $g_{pp}$
in the quasiparticle random-phase approximation (QRPA) model are obtained from
single $\beta$-decay and two-neutrino double beta decay measurements~\cite{Suhonen:2017krv}.
Measurements of the 2$\nu$ECEC half-lives with various nuclei would
shed new light on constraining these parameters.

Natural xenon contains $^{124}$Xe (abundance 0.095\%) and $^{126}$Xe (0.089\%),
in which ECEC can be observed.
$^{124}$Xe has the highest $Q$-value among all the known candidate nuclei for
ECEC at 2864~keV~\cite{AME2012}.
This $Q$-value is sufficiently large to open the $\beta^+$EC and $\beta^+ \beta^+$ channels.
The predictions in the literature for the half-lives of $^{124}$Xe 2$\nu$ECEC
are spread over a wide range between $10^{20}$ and $10^{24}$~years~\cite{Hirsch:1994es,Aunola:1996ui,Rumyantsev:1998uy,Shukla:2007ju,Singh:2007jh,Suhonen:2013rca}
depending on the models used for calculating the corresponding nuclear matrix element and
the effective value of the nucleus' $g_A$.
Although $^{126}$Xe can also undergo 2$\nu$ECEC,
the life-time of this process for $^{126}$Xe is expected to be
much longer than that for $^{124}$Xe
since its $Q$-value is smaller at 920~keV~\cite{AME2012}.

Previous experimental searches for 2$\nu$ECEC on $^{124}$Xe have sought
the simultaneous capture of two $K$-shell electrons (2$\nu$2K) 
using a gas proportional counter with enriched xenon and large-volume liquid xenon (LXe) detectors
with natural xenon as the target.
An experiment with a proportional counter containing 58.6~g of $^{124}$Xe (enriched to 23\%)
published the latest lower bound on the half-life,
$T_{1/2}^{2\nu2K} \left (^{124}{\rm Xe} \right ) > 2.0\times 10^{21}$~years
at 90\% confidence level (CL)~\cite{Gavrilyuk:2014dqa,Gavrilyuk:2015ada}.
Large-volume LXe detectors can also 
observe 2$\nu$ECEC on $^{124}$Xe~\cite{Mei:2013cla,Barros:2014exa}.
The XMASS experiment has conducted a search
with a fiducial xenon mass of 41~kg (containing 39~g of $^{124}$Xe) and
set the most stringent lower limit of
$T_{1/2}^{2\nu2K} \left (^{124}{\rm Xe} \right ) > 4.7\times 10^{21}$~years~\cite{xmass-2nuECEC1}.
The XENON100 experiment also published a result obtained
with a fiducial xenon mass of 34~kg (containing about 29~g of $^{124}$Xe) and
set a lower limit of
$T_{1/2}^{2\nu2K} \left (^{124}{\rm Xe} \right ) > 6.5\times 10^{20}$~years~\cite{Aprile:2016qsw}.
These searches were conducted with similar amount of $^{124}$Xe and live time
as summarized in Table~\ref{table:experimental_results}.
In addition, the XMASS experiment set the first experimental lower limit on
the $^{126}$Xe 2$\nu$2K half-life at
$T_{1/2}^{2\nu2K} \left (^{126}{\rm Xe} \right ) > 4.3\times 10^{21}$~years
using the same data set.

In this paper, we report the results of an improved search for  $^{124}$Xe and $^{126}$Xe
2$\nu$2K events, using data from the XMASS-I detector.
We analyze a new data set taken between November 2013 and July 2016.
The total live time amounts to 800.0~days and the fiducial xenon mass was enlarged to 327~kg
(containing about 311~g of $^{124}$Xe).
We developed a novel method for discriminating the 2$\nu$2K signal from the $\beta$-ray background
using LXe scintillation time profiles.

\begin{table}[tbp]
 \caption{Summary of experimental searches for two-neutrino double electron capture
           on $^{124}$Xe reported to date compared with this work.}
 \label{table:experimental_results}
 \begin{center}
  \begin{tabular}{lccc}
    \hline \hline
    Experiment            & $^{124}$Xe target mass (g) & live time & $T_{1/2}^{2\nu2K} \left (^{124}{\rm Xe} \right )$~($10^{21}$~years)  \\
    \hline
    XMASS (This work)                               & 311  & 800.0~days & $> 21$ \\
    \hline
    XMASS~\cite{xmass-2nuECEC1}                     & 39   & 132.0~days & $> 4.7$ \\   
    Gavrilyuk {\it et al.}~\cite{Gavrilyuk:2015ada} & 58.6 & 3220~h    & $> 2.0$ \\
    XENON100~\cite{Aprile:2016qsw}                  & 29   & 224.6~days & $> 0.65$ \\
    \hline \hline
  \end{tabular}
 \end{center}
\end{table}

\section{The XMASS-I detector}
XMASS-I is a large single-phase LXe detector located
underground (2700~m water equivalent) at the Kamioka Observatory in Japan~\cite{xmass-detector}. 
An active target of 832~kg of LXe is held inside a
pentakis-dodecahedral copper structure that hosts 642 inward-looking 2-inch
Hamamatsu R10789 photomultiplier tubes (PMTs) on its approximately spherical inner surface
at a radius of about 40~cm.
The photocathode coverage of the inner surface is 62.4\%.
Signals from each PMT are recorded with CAEN V1751 waveform digitizers
with a sampling rate of 1~GHz and 10-bit resolution.

The gains of the PMTs are monitored weekly using a blue LED embedded in the inner surface of the detector. 
The scintillation yield response is traced with a $^{57}$Co source~\cite{xmass-source}
inserted along the central vertical axis of the detector every week or two. 
Through measurements with the $^{57}$Co source at the center of the detector volume, 
the photoelectron (PE) yield was determined to be $\sim$15~PE/keV for 122~keV $\gamma$-rays.
The nonlinear response of the scintillation yield for electron-mediated events in the detector
was calibrated over the energy range from 5.9~keV to 2614~keV
with $^{55}$Fe, $^{241}$Am, $^{109}$Cd, $^{57}$Co, $^{137}$Cs, $^{60}$Co, and $^{232}$Th sources.
Hereinafter, this calibrated energy is represented as keV$_{\rm ee}$
where the subscript stands for the electron-equivalent energy.
The timing offsets for the PMT channels owing to the differences in their cable lengths and
the electronic responses were also traced by the $^{57}$Co calibration.

The LXe detector is located at the center of a cylindrical water Cherenkov detector,
which is 11~m in height and 10~m in diameter.
The outer detector is equipped with 72 20-inch Hamamatsu H3600 PMTs.
This detector acts as an active veto counter for cosmic-ray muons as well as
a passive shield against neutrons and $\gamma$-rays from the surrounding rock.

Data acquisition is triggered if at least four inner-detector PMTs record
a signal within 200~ns
or if at least eight outer-detector PMTs register a signal within 200~ns.
A 50~MHz clock is used to measure the time difference between triggers.
One-pulse-per-second (1PPS) signals from the global positioning system (GPS) are fed as triggers
for precise time stamping.
The GPS 1PPS triggers are also used to flash the LED for the PMT gain monitoring.

\section{Expected signal and simulation}
The process of 2$\nu$ECEC on $^{124}$Xe is
\begin{equation}
    ^{124}{\rm Xe} + 2e^- \to ^{124}{\rm Te} + 2\nu_e \ .
\end{equation}
If two $K$-shell electrons in the $^{124}$Xe atom are captured simultaneously,
a daughter atom of $^{124}$Te is formed with two vacancies in the $K$-shell and
this atom relaxes by emitting atomic $X$-rays and/or Auger electrons.
Our Monte Carlo simulations of the atomic de-excitation signal are based on
the atomic relaxation package in Geant4 \cite{2007ITNS...54..585G}.
On the assumption that the $X$-rays and Auger electrons emitted in the 2$\nu$2K event
are like those generated by two single $K$-shell vacancies,
the signal simulation begins with two Te atoms with a single $K$-shell vacancy.
In such a case, the total energy deposition is given by
twice the $K$-shell binding energy of Te ($2K_{b} = 63.63$~keV).
On the other hand, the energy of the two electron holes in the $K$-shell of $^{124}$Te
is calculated to be 64.46~keV~\cite{Nesterenko:2012xp}, which only varies by 0.8~keV.
Since the energy resolution of the 2$\nu$2K signal peak is estimated to be 3.2~keV
after all the detector responses mentioned below are accounted for,
we judge that this difference is negligible in our analysis.
The results actually do not change even if the peak position of the simulated signal
is artificially shifted by this amount.
According to the simulation, 77\% of 2$\nu$2K events emit two $K$-shell $X$-rays,
21\% of events emit a single $K$-shell $X$-ray, and the remaining 1.6\% of events emits
no $K$-shell $X$-ray.
These probabilities are consistent with those expected from the fluorescence yield
for the $K$-shell of Te, $\omega_K=0.875$~\cite{TOI}.
Auger electron cascades are also simulated. 
The energy deposition from the recoil of the daughter nucleus is $\sim$30~eV
at most, which is negligible.
Simulated de-excitation events are generated uniformly throughout the detector volume.

The nonlinearity of scintillation yield is accounted for
using the nonlinearity model from Doke et al.~\cite{LXeLightYield}
with a further correction obtained from the $\gamma$-ray calibrations.
The absolute energy scale of the simulation is adjusted at 122~keV.
The time profile of the scintillation is also modeled based on
the $\gamma$-ray calibrations~\cite{xmass-decaytime}.
Propagation of scintillation photons in LXe is also simulated.
Optical parameters of the LXe such as absorption and scattering lengths
for the scintillation are tuned by source calibration data at various positions.
The group velocity of the scintillation light in the LXe follows from
LXe's refractive index ($\sim$11~cm/ns for 175-nm light~\cite{LXeSpeed}).
Charge and timing responses of the PMTs are also modeled in the simulation
based on the calibrations with the LED and the $\gamma$-ray sources.
Finally, waveforms of the PMT signal are simulated using the template of a single-PE
waveform obtained from the LED calibration data.

\section{Data set}
The data used in the present analysis were collected between November 20, 2013 and July 20, 2016.
The data set was divided into four periods depending on the detector
conditions at that time as summarized
in Table~\ref{table:data-set}.
Period 1 started two weeks after the introduction of LXe into the detector.
At the beginning of the run, we observed neutron-activated peaks from
$^{131\rm m}$Xe and $^{129\rm m}$Xe that were created when the LXe was stored outside the water shield.
We also performed the $^{252}$Cf calibration data collection twice in this period.
Runs within 10~days after each calibration were excluded from the data set.
We ended period 1 60~days after the second $^{252}$Cf calibration
because the $^{131\rm m}$Xe and $^{129\rm m}$Xe peaks caused by the $^{252}$Cf disappeared.
Period 2 then ran until the continuous gas circulation
at a flow rate of $\sim$1.5~L/min with a getter purifier was introduced.
Period 3 was ended so that the xenon could be purified
by vaporizing xenon once to remove possible non-volatile impurities dissolved in LXe.
During the purification process, LXe was extracted from the detector, and therefore,
xenon was exposed to and activated by thermal neutrons outside the water shield.
The purification process took 7 days and period 4 started immediately after
completing the introduction of LXe into detector.

We selected periods of operation under what we call normal data taking conditions
with a stable temperature (172.6--173.0~K) and pressure (0.162--0.164~MPa absolute) of the LXe in the detector.
After further removing periods of operation that include excessive PMT noise,
unstable pedestal levels, or abnormal trigger rates, the total live time became 800.0~days.

\begin{table}[tbp]
 \caption{Summary of the data set used in this analysis.}
 \label{table:data-set}
 \begin{center}
  \begin{tabular}{lcccl}
    \hline \hline
    Period & Start date$-$End date & Live time (days) & Gas circulation & Comment \\
    \hline
    1 & Nov 20, 2013$-$May 13, 2014 & 124.0 & None            & Activated \\
    2 & May 13, 2014$-$Mar 13, 2015 & 249.1 & None            & \\
    3 & Mar 13, 2015$-$Mar 29, 2016 & 338.1 & $\sim$1.5~L/min & \\
    4 & Apr 14, 2016$-$Jul 20, 2016 &  88.8 & $\sim$1.5~L/min & Activated \\
    \hline \hline
  \end{tabular}
 \end{center}
\end{table}

\section{Event reduction and classification}
The event-reduction process comprises four steps: pre-selection, the fiducial volume selection,
$^{214}${\rm Bi} identification, and particle identification.

\subsection{Pre-selection}
Pre-selection requires that no outer-detector trigger is associated with an event,
that the time elapsed since the previous inner-detector event ($dT_{\rm pre}$)
is at least 10~ms,
and that the standard deviation of the inner-detector hit timing distribution in the event is less than 100~ns.
The last two requirements remove events caused by after-pulses in the PMTs following bright events.
The $dT_{\rm pre}$ cut eliminates events by chance coincidence at a probability of 3.0\% on average,
which was estimated from the fraction of the GPS 1PPS events that are rejected by this cut.
The chance coincidence probability is counted as dead time, and the live time mentioned above is obtained
after subtracting this dead time.

\subsection{Fiducial volume selection}
To select events that occurred within the fiducial volume, an event vertex is reconstructed
based on a maximum-likelihood evaluation of the observed light distribution in the detector~\cite{xmass-detector}.
We select events whose reconstructed vertex has a radial distance of
less than 30~cm from the center of the detector.
The fiducial mass of natural xenon in that volume is 327~kg,
containing 311~g of $^{124}$Xe and 291~g of $^{126}$Xe.

\subsection{$^{214}${\rm Bi} identification}
$^{222}$Rn emanates from the detector's surface and contaminates the LXe within the detector.
Thus, its daughters, $^{214}$Bi and $^{214}$Pb, become one of the major sources of
$\beta$-ray background.
$^{214}$Bi can be tagged using the $^{214}$Bi-$^{214}$Po delayed coincidence ($T_{1/2}=164$~$\mu$s)
and is used as a good control sample of pure $\beta$-ray events in the relevant energy range.
To remove the $^{214}$Bi events,
events whose time difference from the subsequent event ($dT_{\rm post}$)
is less than 1~ms are rejected from the 2$\nu$2K signal sample.
This cut reduces the $^{214}$Bi background by a factor of $\sim$70,
while consequently discarding only 0.4\% of all other events.
The counterpart sample, \textit{i.e.} events with $0.015~{\rm ms}< dT_{\rm post}< 1$~ms,
is referred to as the $^{214}$Bi sample and is used to constrain the $^{214}$Bi and $^{214}$Pb backgrounds.

\subsection{Particle identification}
The scintillation time profiles of LXe can be used for particle identification.
We use them to eliminate
both the $\alpha$-ray and $\beta$-ray backgrounds from the 2$\nu$2K signal sample.

After the fiducial volume selection, the largest source of background in the relevant energy range
are $\beta$-rays coming from radioactive impurities within the LXe.
2$\nu$2K or $\gamma$-ray events can be discriminated from these $\beta$-ray events
by utilizing the energy dependence of the scintillation decay time for electron-induced events.
The scintillation decay time increases from 28~ns to 48~ns
as the kinetic energy of an electron increases from 3~keV to 1~MeV
as summarized in Fig.~3 of \cite{xmass-decaytime}.
In the case of the 2$\nu$2K or $\gamma$-ray events, the $X$-ray or $\gamma$-ray is converted into
multiple low-energy electrons in the LXe; this shortens the effective scintillation decay time by
a few ns from that of an event caused by a single electron with the same deposited energy. 
Especially, events caused by 2$\nu$2K or $\gamma$-rays with energy close to twice of the
$K$-shell binding energy are easily-distinguishable from the $\beta$-ray events
by this effective scintillation decay time.

\begin{figure}[tbp]
  \begin{center}
    \includegraphics[keepaspectratio=true,height=85mm]{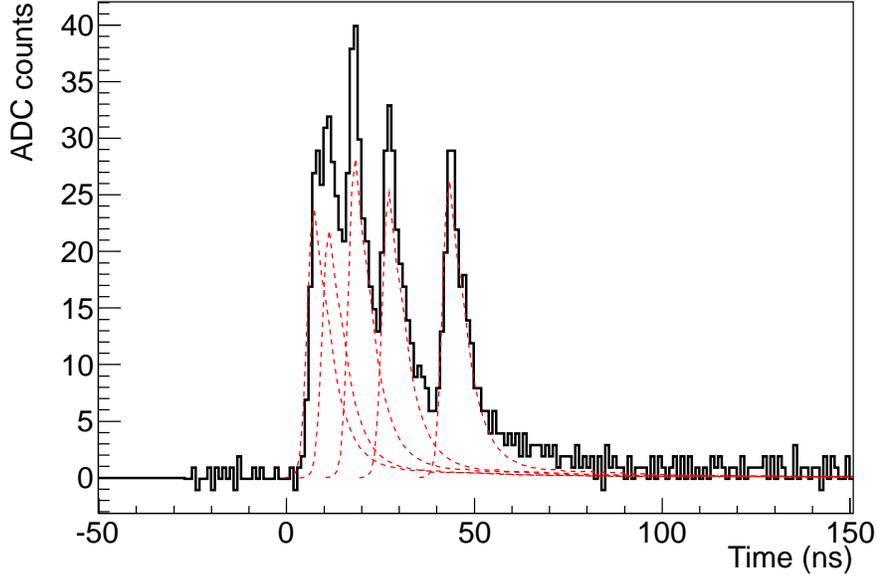}
  \end{center}
  \caption{An example waveform recorded in a PMT from the $^{241}$Am calibration data.
  The $^{241}$Am source was placed at $z=20$~cm.
  Red dashed curves represent decomposed single-PE pulses.}
  \label{fig:waveform.eps}
\end{figure}

The particle-identification parameter $\beta$CL is formulated as follows.
First, waveforms in each PMT are decomposed into single-PE pulses using the single-PE waveform template~\cite{xmass-decaytime}.
Figure~\ref{fig:waveform.eps} shows an example of a waveform recorded in a PMT from the $^{241}$Am calibration data.
Decomposed single-PE pulses are also shown in the figure.
The timings of the decomposed single-PE pulses in all the PMTs are sorted chronologically 
after correcting for the time-of-flight of scintillation photons.
The single-PE pulses in the first 20~ns are excluded from the following calculation
to avoid systematic uncertainties in the leading edge of the scintillation time profile.
The variable $\beta$CL is defined as
\begin{equation}
  \beta {\rm CL} = P \times \sum_{i=0}^{n-1} \frac{(-\ln P)^{i}}{i!} \ ,
\end{equation}
where $n$ is the total number of single-PE pluses after truncating the first 20~ns, $P = \prod_{i=0}^{n-1} {\rm CL}_i$, and
CL$_i$ ($i$=0, 1, 2, \ldots, $n-1$) is the CL of each pulse timing under the assumption
that the event is caused by a $\beta$-ray.
The probability-density function of the pulse-timing distribution for a $\beta$-ray event
including its energy and position dependences is modeled from measurements in
the $^{214}$Bi data sample over the energy range between 30 and 200~keV$_{\rm ee}$.
This formula is in general used to combine $p$-values from a set of independent tests
for a certain hypothesis~\cite{Fisher}.

\begin{figure}[tbp]
  \begin{minipage}{0.5\hsize}
    \begin{center}
      \includegraphics[keepaspectratio=true,width=80mm]{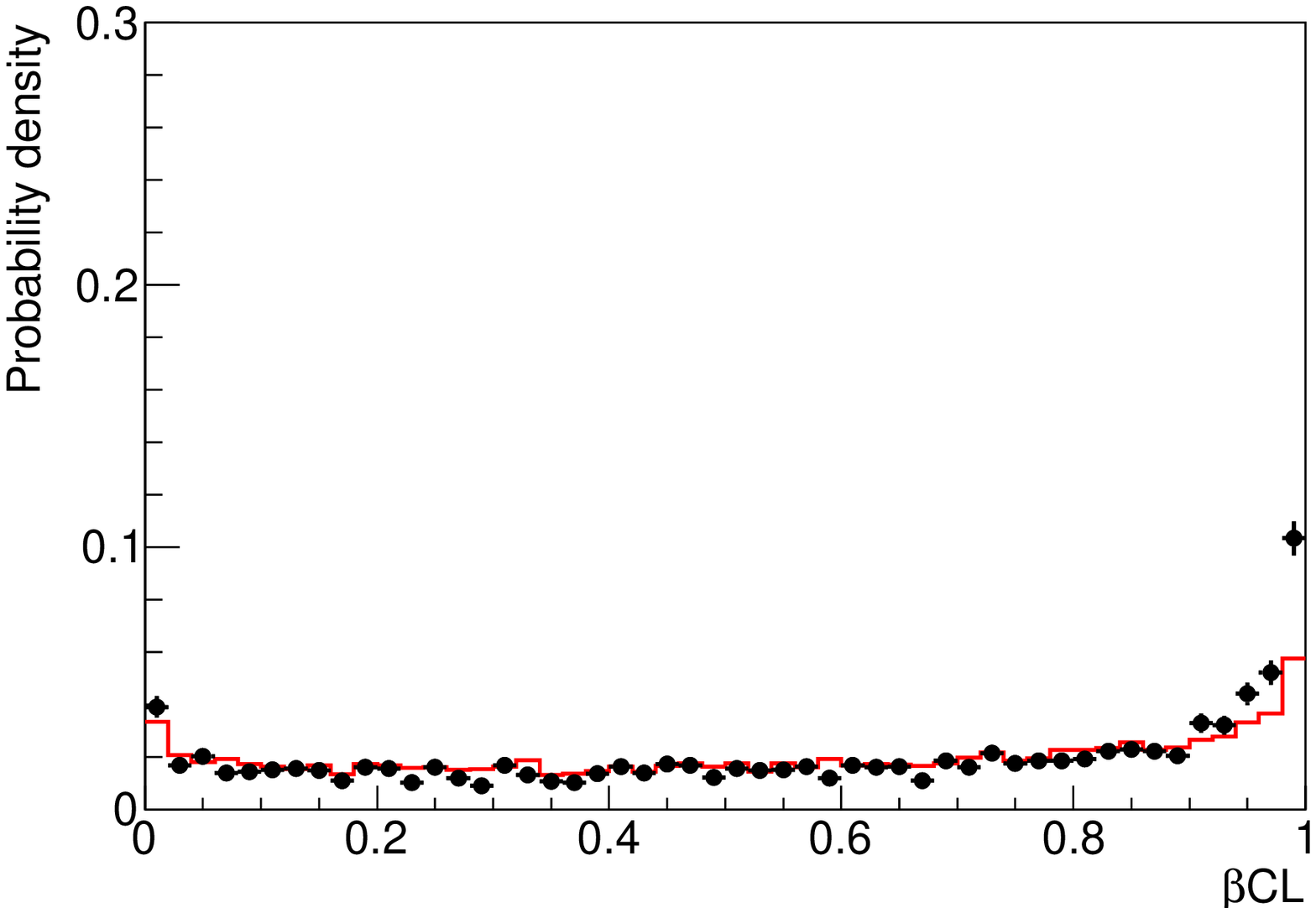}
    \end{center}
  \end{minipage}
  \begin{minipage}{0.5\hsize}
    \begin{center}
      \includegraphics[keepaspectratio=true,width=80mm]{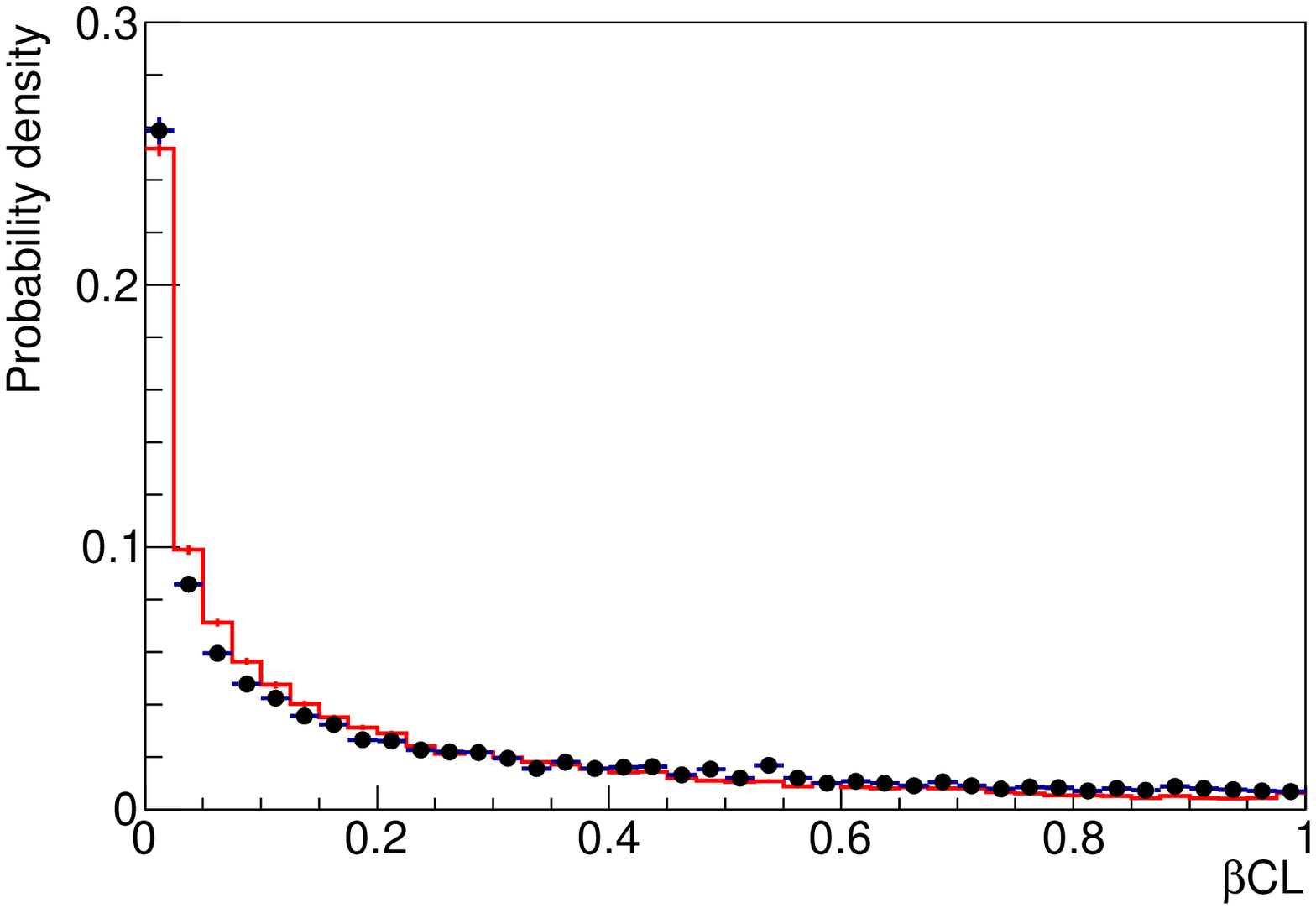}
    \end{center}
  \end{minipage}
  \caption{The particle-identification parameter $\beta$CL for the $^{214}$Bi $\beta$-ray events
  in the energy range from 30 to 200~keV$_{\rm ee}$ (left) and the $^{241}$Am 59.5~keV $\gamma$-ray events (right).
  The distributions of the observed data (black points) and the simulated events (red curves) are normalized to the unit area.}
  \label{fig:betacl.eps}
\end{figure}

Figure~\ref{fig:betacl.eps} shows distributions of the variable $\beta$CL for the $^{214}$Bi sample
in the energy range from 30 to 200~keV$_{\rm ee}$
along with the $^{241}$Am 59.5~keV $\gamma$-ray events. 
While the $\beta$-ray events in the $^{214}$Bi sample are distributed between 0 and 1,
the distribution of the 59.5~keV $\gamma$-ray events in the $^{241}$Am sample peaks at $\beta$CL = 0. 
Events with $\beta$CL less than 0.05 are classified as the $\beta$-depleted sample,
and the rest is referred to as the $\beta$-enriched sample.
When selecting events with $\beta$CL less than 0.05, 42\% of the 2$\nu$2K signal events are selected,
while only 6\% of the $\beta$-ray events from the $^{214}$Bi decay in this energy range 
are selected. 
Thus, the signal-to-noise ratio is improved by a factor of 7 by this selection. 
The cut position is tuned based on the simulated data to maximize the sensitivity for
the 2$\nu$2K signal.

$\alpha$-ray events often occur in the grooves of the inner surface of the detector, so that
only some of their energy is detected. These events are sometimes incorrectly reconstructed
within the fiducial volume~\cite{xmass-fiducial}. 
Above 30~keV$_{\rm ee}$, $\alpha$-ray events can be clearly separated from $\beta$-ray or $\gamma$-ray events
using the scintillation decay time.
Therefore, the waveforms from all PMTs are summed up to form a total waveform of the event
after correcting for the relative gain and timing of each PMT.
Then, the falling edge of that total waveform is fitted with an exponential function to obtain the decay time for each event.
Events with fitted decay times of less than 30~ns are deemed $\alpha$-ray events and are rejected.

\begin{figure}[tbp]
  \begin{minipage}{0.5\hsize}
    \begin{center}
      \includegraphics[keepaspectratio=true,width=80mm]{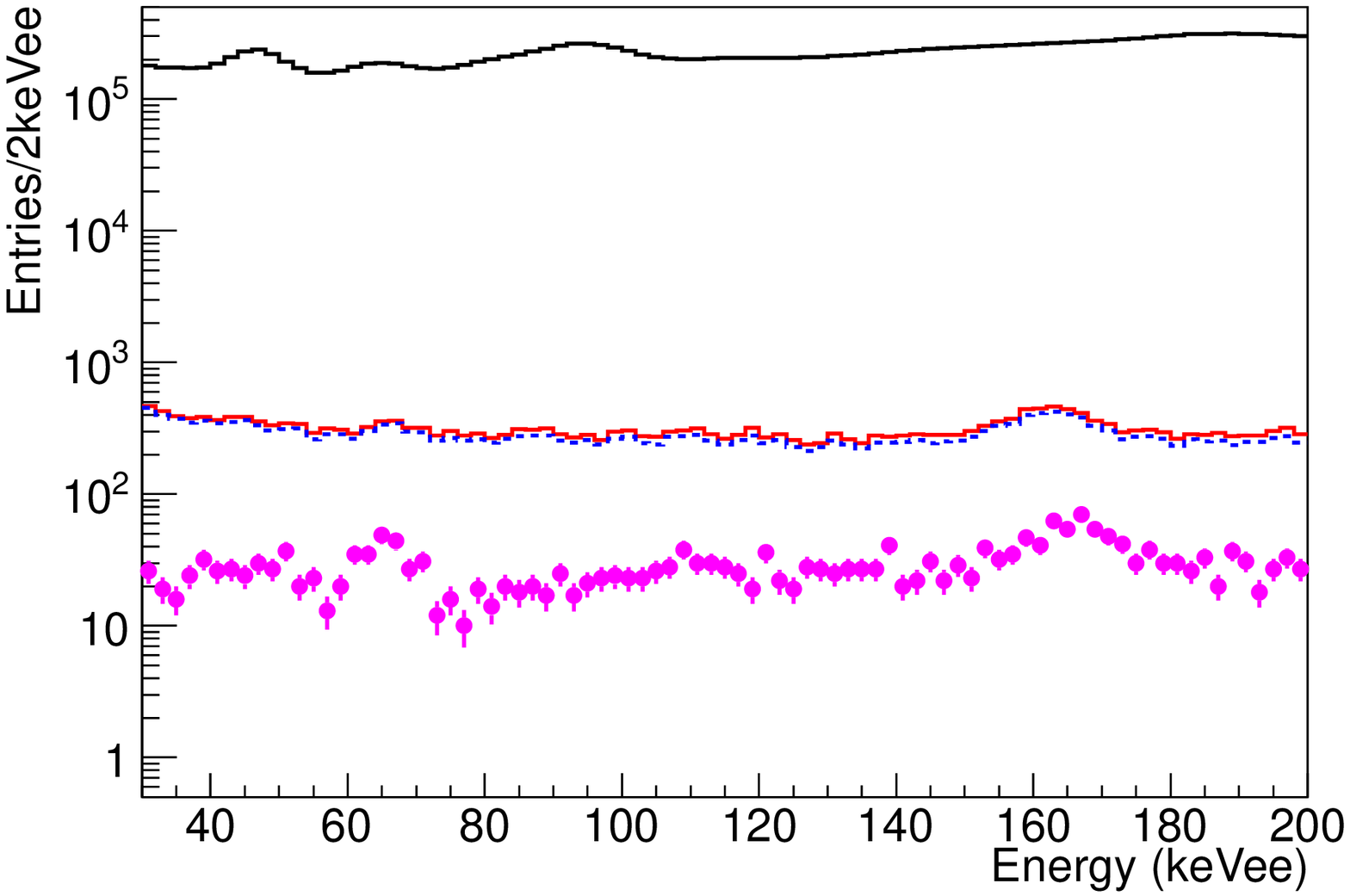}
    \end{center}
  \end{minipage}
  \begin{minipage}{0.5\hsize}
    \begin{center}
      \includegraphics[keepaspectratio=true,width=80mm]{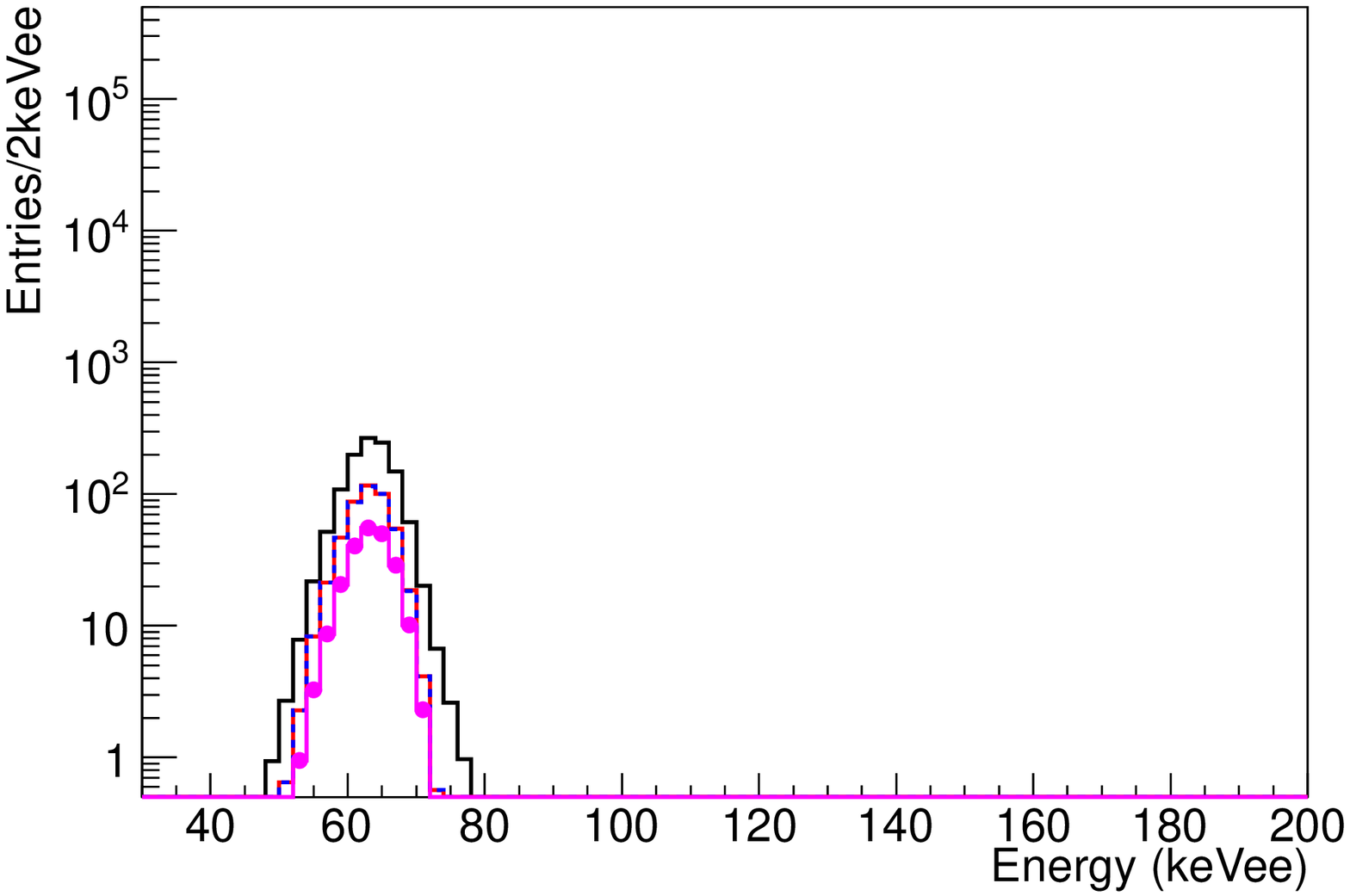}
    \end{center}
  \end{minipage}
  \caption{Energy spectra after each event reduction step for the observed data (left) and
  the simulated 2$\nu$2K sample assuming $T_{1/2}=4.7\times 10^{21}$~years (right).
  From top to bottom, energy distributions after the pre-selection (black solid),
  fiducial volume selection (red solid), $^{214}$Bi rejection (blue dashed
  nearly overlaps with red solid),
  and $\beta$-like event rejection (magenta points) steps are shown.}
  \label{fig:data_dec_spectrum.eps}
\end{figure}

Figure~\ref{fig:data_dec_spectrum.eps} shows the energy spectra plotted
after each event-reduction step for the observed data and the simulated 2$\nu$2K sample.
For the simulated 2$\nu$2K sample, $T_{1/2}=4.7\times 10^{21}$~years is assumed.

\section{Spectrum fitting}
\subsection{Chi-square definition}
To extract the 2$\nu$2K signal from the observed data, the energy spectra
for the $\beta$-depleted samples, $\beta$-enriched samples, and $^{214}$Bi samples
are simultaneously fitted to the expected signal and background spectra.
The energy range from 30 to 200~keV$_{\rm ee}$ is used for fitting.
The chi-square value is defined as
\begin{eqnarray}
  \chi^2 &=& -2\ln L \nonumber \\
         &=& 2\sum_{i=1}^{N_{\rm sample}} \sum_{j=1}^{N_{\rm period}} \sum_{k=1}^{N_{\rm bin}}
  \left ( n_{ijk}^{\rm exp}(p_l)-n_{ijk}^{\rm data}+n_{ijk}^{\rm data} \ln \frac{n_{ijk}^{\rm data}}{n_{ijk}^{\rm exp}(p_l)} \right )
  + \sum_{l=1}^{N_{\rm sys}} \frac{(1-p_l)^2}{\sigma_l^2} \ ,
\end{eqnarray}
where $n_{ijk}^{\rm data}$, and $n_{ijk}^{\rm exp} (p_l)$ are the observed and expected number of events
in $i$-th sub-sample and $j$-th period and $k$-th energy bin, respectively.
$N_{\rm sample}=3$, $N_{\rm period}=4$, $N_{\rm bin}=85$, and $N_{\rm sys}$ are the number of sub-samples,
periods, energy bins, and constrained systematic parameters, respectively.
$p_l$ and $\sigma_l$ are a scaling parameter for the nominal value and its relative error, respectively.

\subsection{Expected background}
We consider three types of backgrounds: radioactive isotopes (RIs) in the LXe,
neutron activation of xenon, and external backgrounds. 

For the internal RIs, the $^{222}$Rn daughters ($^{214}$Bi and $^{214}$Pb), $^{85}$Kr, $^{39}$Ar, $^{14}$C,
and $^{136}$Xe are considered.
The $^{214}$Bi activity during each period is determined from the fitting to the $^{214}$Bi sample
and the $^{214}$Pb activity in each period follows from this $^{214}$Bi activity
as both originate from $^{222}$Rn.
While $^{85}$Kr decays by $\beta$-decay ($Q_\beta=687$~keV, $T_{1/2}=10.8$~years)
predominantly into the ground state of $^{85}$Rb,
0.434\% of their decays go into the 514-keV excited state of $^{85}$Rb
followed by a nuclear relaxation $\gamma$-ray ($T_{1/2}=1.014~\mu$s).
$^{85}$Kr contamination in the detector is measured to be $0.26\pm 0.06$~mBq
by the coincidence of $\beta$-ray and $\gamma$-ray events.
In this analysis, the $^{85}$Kr activity in each period is fitted with this constraint.
We have also found argon contamination in the xenon through measurements of the sampled xenon gas
using gas chromatography-mass spectrometry (GC-MS).
The argon is thought to have adsorbed to the detector material
when we conducted a leakage test of the LXe chamber using argon gas in 2013.
$^{39}$Ar undergoes $\beta$-decay ($Q_\beta=565$~keV, $T_{1/2}=269$~years).
By comparing the energy spectra for periods 2 and 3 of the data set,
we found a reduction in event rate below $\sim$150~keV$_{\rm ee}$
in the $\beta$-enriched sample.
The difference in the energy spectra between two periods is consistent with
the $\beta$-decay of $^{14}$C ($Q_\beta=156$~keV, $T_{1/2}=5730$~years).
Hence, we assume that impurities containing carbon were reduced by
gas circulation through the getter although its chemical form is not known.
Finally, natural xenon contains $^{136}$Xe with an isotopic abundance of 8.9\%,
and $^{136}$Xe undergoes 2$\nu \beta \beta$ decay
($Q_{\beta \beta}=2.46$~MeV, $T_{1/2}=2.2\times 10^{21}$~years~\cite{Albert:2013gpz,KamLAND-Zen:2016pfg}).

Although the LXe detector is shielded against environmental neutrons by water,
some of the detector components
such as the cable feed-through box, calibration system, and cryogenic system lie
outside the water shield and are filled with xenon gas~\cite{xmass-detector}.
The volume of xenon gas outside the water shield is estimated to be $2.6\times 10^5$~cm$^3$
at the standard temperature of 273.15~K and pressure of $10^5$~Pa.
This xenon is activated by thermal neutron capture and returned to the LXe in the detector.
The resulting 13 RIs, $^{125}$Xe, $^{125\rm m}$Xe, $^{125}$I, $^{127}$Xe, $^{127\rm m}$Xe,
$^{129\rm m}$Xe, $^{131\rm m}$Xe, $^{133}$Xe, $^{133\rm m}$Xe, $^{135}$Xe, $^{135\rm m}$Xe, $^{137}$Xe, and $^{137}$Cs, are also considered.
Their activities are calculated based on the isotopic abundance of xenon and
the cross sections of thermal neutron capture.
Among those isotopes, $^{125}$I is the most considerable background in this analysis.
The $^{125}$I is produced from $^{125}$Xe and $^{125\rm m}$Xe created by
thermal neutron capture on $^{124}$Xe with a total cross section of
165$\pm$11~barn~\cite{neutron-capture}.
$^{125}$I decays by 100\% electron capture via an excited state of $^{125}$Te
into the ground state of $^{125}$Te with a total energy deposition of 67.5~keV.
The flux of thermal neutrons ($E<0.5$~eV) in the Kamioka mine has been measured to be
(0.8-1.4)$\times 10^{-5}$~/cm$^2$/s~\cite{neutron-flux1,neutron-flux2}.
In this analysis, the thermal neutron flux during each period is fitted
under the constraint of these measurements.
$^{125}$I, $^{131\rm m}$Xe, and $^{133}$Xe are the main RIs relevant to this analysis and
the entire energy range of the beta-depleted samples has the power
to constrain the thermal neutron flux.
Activations of xenon by neutrons emitted from the $(\alpha,n)$ reaction or
spontaneous fission in the detector material is negligible.
In addition, occasional neutron activations of the LXe appear in periods 1 and 4
due to the $^{252}$Cf calibration and the purification works.
The data taken just after the $^{252}$Cf calibration, which were excluded from
this analysis, showed clear event rate increases in the energy range between 30 and 200 keV
due to $^{131\rm m}$Xe and $^{133}$Xe.
To accommodate these backgrounds, we introduce additional quantities of
$^{131\rm m}$Xe and $^{133}$Xe in the fitting.

For the external backgrounds, a detailed evaluation of radioactive backgrounds
from each detector material has been conducted previously~\cite{xmass-fiducial}.
In the present data set, a small contribution of $\gamma$-ray backgrounds
from impurities in the PMTs is expected.
$^{238}$U, $^{232}$Th, $^{60}$Co, and $^{40}$K are considered,
and the uncertainties in their activities are accounted for in the fitting.

\begin{table}[tbp]
 \caption{Summary of systematic parameters and their uncertainties used as constraints
           in the spectrum fitting.}
 \label{table:systematic_parameters}         
 \begin{center}
  \begin{tabular}{lccc}
    \hline \hline
    Item      & Fractional uncertainty & Period dependence \\
    \hline
    $^{222}$Rn daughters ($^{214}$Bi and $^{214}$Pb) & Unconstrained & Assumed \\
    $^{85}$Kr & $\pm$23\% & Assumed \\
    $^{39}$Ar & Unconstrained & Assumed \\
    $^{14}$C  & Unconstrained & Assumed \\
    Thermal neutron flux & $\pm$27\% & Assumed\\
    Additional $^{131\rm m}$Xe  & Unconstrained & Assumed \\
    Additional $^{133}$Xe       & Unconstrained & Assumed \\
    $\gamma$-ray backgrounds from PMTs & $\pm$9.4\% ($^{238}$U), $\pm$24\% ($^{232}$Th), & Not assumed \\
                              & $\pm$11\% ($^{60}$Co), $\pm$17\% ($^{40}$K) \\
    \hline
    Isotopic abundance    & $\pm$8.5\% ($^{124}$Xe), $\pm$12\% ($^{126}$Xe) & Not assumed \\    
    Fiducial volume       & $\pm$4.5\%    & Not assumed   \\
    $\beta$CL acceptance for $\gamma$-ray & $\pm$30\% & Not assumed \\
    $\beta$CL acceptance for $\beta$-ray  & $\pm$8.0\% & Not assumed \\
    Energy scale ($\beta$-depleted sample) & $\pm$2.0\% & Assumed \\
    Energy scale ($\beta$-enriched sample) & $\pm$2.0\% & Assumed \\
    \hline \hline
  \end{tabular}
 \end{center}
\end{table}

\subsection{Systematic uncertainties}
Systematic uncertainties in the background yields, exposure, event selections, and energy scales 
are considered in the fitting
as listed in Table~\ref{table:systematic_parameters}.
The upper part of Table 3 summarizes the systematic parameters used to determine the
activities of RI backgrounds in the spectrum fitting.
The $^{85}$Kr activity, thermal neutron flux, and $\gamma$-ray backgrounds from the PMTs
are constrained by the external measurements as described in the previous section
since this spectrum fitting does not have the sensitivity for an independent evaluation.

Isotopic composition of the LXe was measured with a mass spectrometer,
and the result was consistent with that of natural xenon in air~\cite{xmass-2nuECEC1}.
The uncertainties in the measurement,
$\pm$8.5\% for $^{124}$Xe and $\pm$12\% for $^{126}$Xe, are treated as a systematic error.
The uncertainties in the LXe density and the detector live time are negligible.
The uncertainties in the event selections and energy scales are estimated from
comparisons between data and simulated samples for the $^{241}$Am (59.5~keV $\gamma$-ray) and $^{57}$Co (122~keV $\gamma$-ray)
calibration data at various positions within the fiducial volume.
The radial position of the reconstructed vertex for the calibration data differs from that for the simulated result
by $\pm 4.5$~mm near the fiducial volume boundary, which causes $\pm 4.5$\% uncertainty in the fiducial LXe volume.
From the difference in the $\beta$CL distribution between the calibration data
and simulated samples, the uncertainty in acceptance of the $\beta$CL cut for
the $\gamma$-ray events is found to be $\pm$30\%.
In the same manner, the uncertainty in the rejection power of the $\beta$CL cut
for $\beta$-ray events is evaluated to be $\pm$8.0\% from the comparison of
the $\beta$CL distributions for the $^{214}$Bi sample
in the energy range from 30 to 200~keV$_{\rm ee}$.
By comparing the peak position of the $\gamma$-ray calibration data and simulated samples
at various source positions and in different periods, the uncertainty in energy scale
for the $\gamma$-ray events is estimated to be $\pm$2.0\%.
Since we observe a small difference in the peak position of the $\gamma$-ray calibration
data between the $\beta$-depleted and $\beta$-enriched samples,
the energy scales for the $\beta$-depleted and $\beta$-enriched samples are treated
independently.

\section{Results and discussion}

\begin{figure}[tbp]
  \begin{center}
    \includegraphics[keepaspectratio=true,width=155mm]{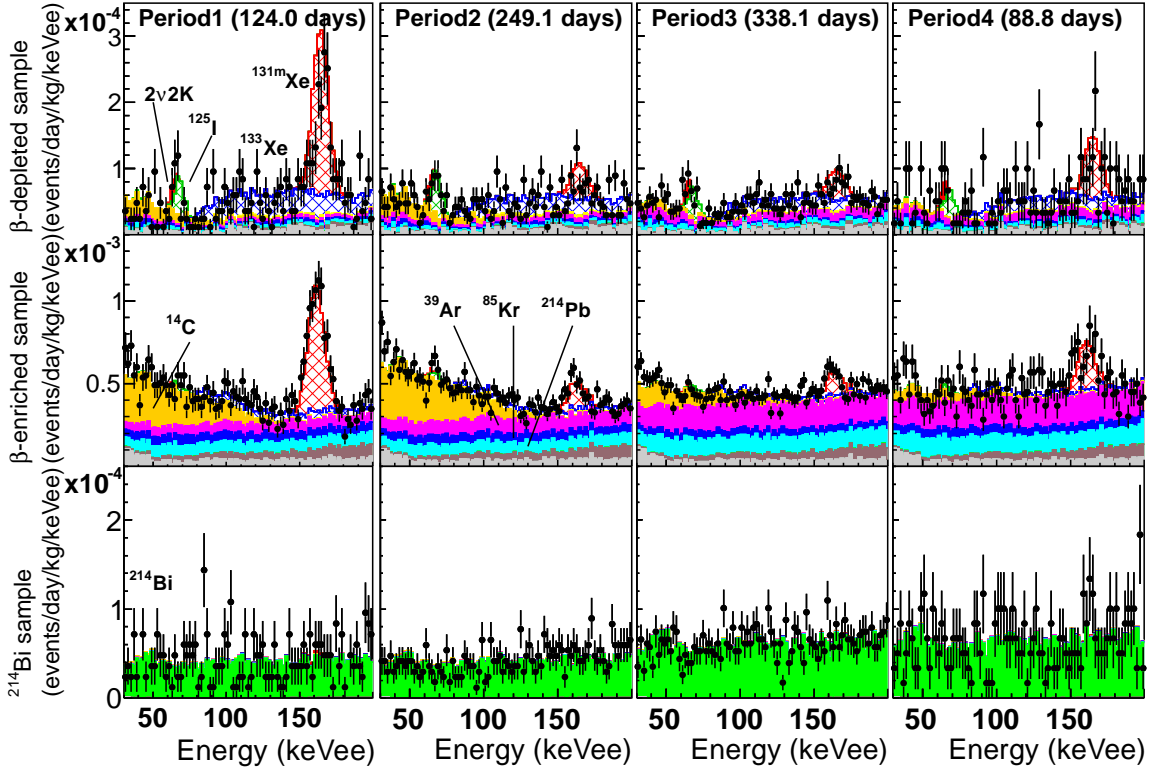}
  \end{center}
  \caption{Energy spectra for the $\beta$-depleted samples (top), $\beta$-enriched samples (middle),
  and $^{214}$Bi samples (bottom). The observed data spectra (points) are overlaid with the best-fit
  2$\nu$2K signal and background spectra (colored stacked histograms).
  Colored histograms are the 2$\nu$2K signal (red filled),
  $^{125}$I (green hatched), $^{131\rm m}$Xe (red hatched), $^{133}$Xe (blue hatched),
  $^{14}$C (orange filled), $^{39}$Ar (magenta filled), $^{85}$Kr (blue filled),
  $^{214}$Pb (cyan filled), $^{214}$Bi (green filled), $^{136}$Xe 2$\nu \beta \beta$ (brown filled),
  and external backgrounds (gray filled).
  }
  \label{fig:fit-result-spectra.eps}
\end{figure}

\begin{figure}[tbp]
  \begin{center}
    \includegraphics[keepaspectratio=true,height=85mm]{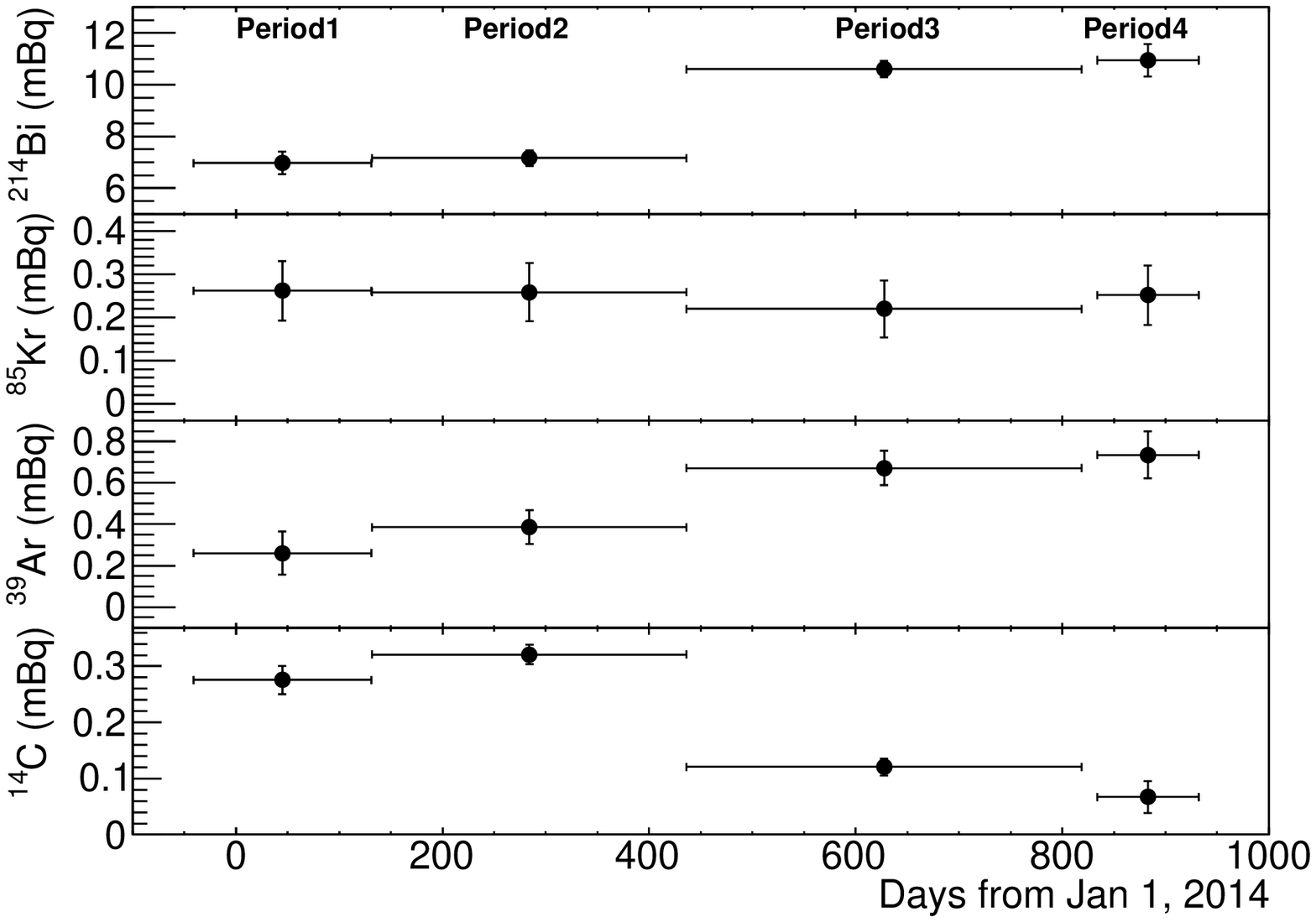}
  \end{center}
  \caption{Time variation of the fitted activities of $^{214}$Bi, $^{85}$Kr, $^{39}$Ar, and $^{14}$C
  in the active 832~kg LXe volume.
  The left and right edges of the horizontal error bars represent the start and end of each period,
  respectively.}
  \label{fig:fit-result-parameters1.eps}
\end{figure}

Figure~\ref{fig:fit-result-spectra.eps} shows the energy spectra for the $\beta$-depleted samples,
$\beta$-enriched samples, and $^{214}$Bi samples.
The observed spectra are overlaid with the best-fit 2$\nu$2K signal and background spectra.
The best fit result gives $\chi^2$/ndf = 1073/999.
The bottom figures determine the activities of $^{214}$Bi in LXe and constrain the $^{214}$Pb
activities.
The middle figures determine the $^{39}$Ar and $^{14}$C activities while the amount of $^{85}$Kr
is constrained by the independent $\beta$-$\gamma$ coincidence measurement.
The variation in time of the fitted activities of $^{214}$Bi, $^{85}$Kr, $^{39}$Ar, and $^{14}$C
in the active 832~kg LXe volume are shown in Fig.~\ref{fig:fit-result-parameters1.eps}.
The $^{222}$Rn concentration in LXe increased by $\sim$50\% after gas circulation
was initiated at the beginning of period 3.
It is surmised that $^{222}$Rn emanating from detector materials in the xenon gas volume
mixes into the LXe by the gas circulation.
An increase of the $^{39}$Ar concentration in period 3 is thought to occur in the same manner.
On the other hand, the $^{14}$C concentration decreased with gas circulation. 

\begin{figure}[tbp]
  \begin{center}
    \includegraphics[keepaspectratio=true,height=85mm]{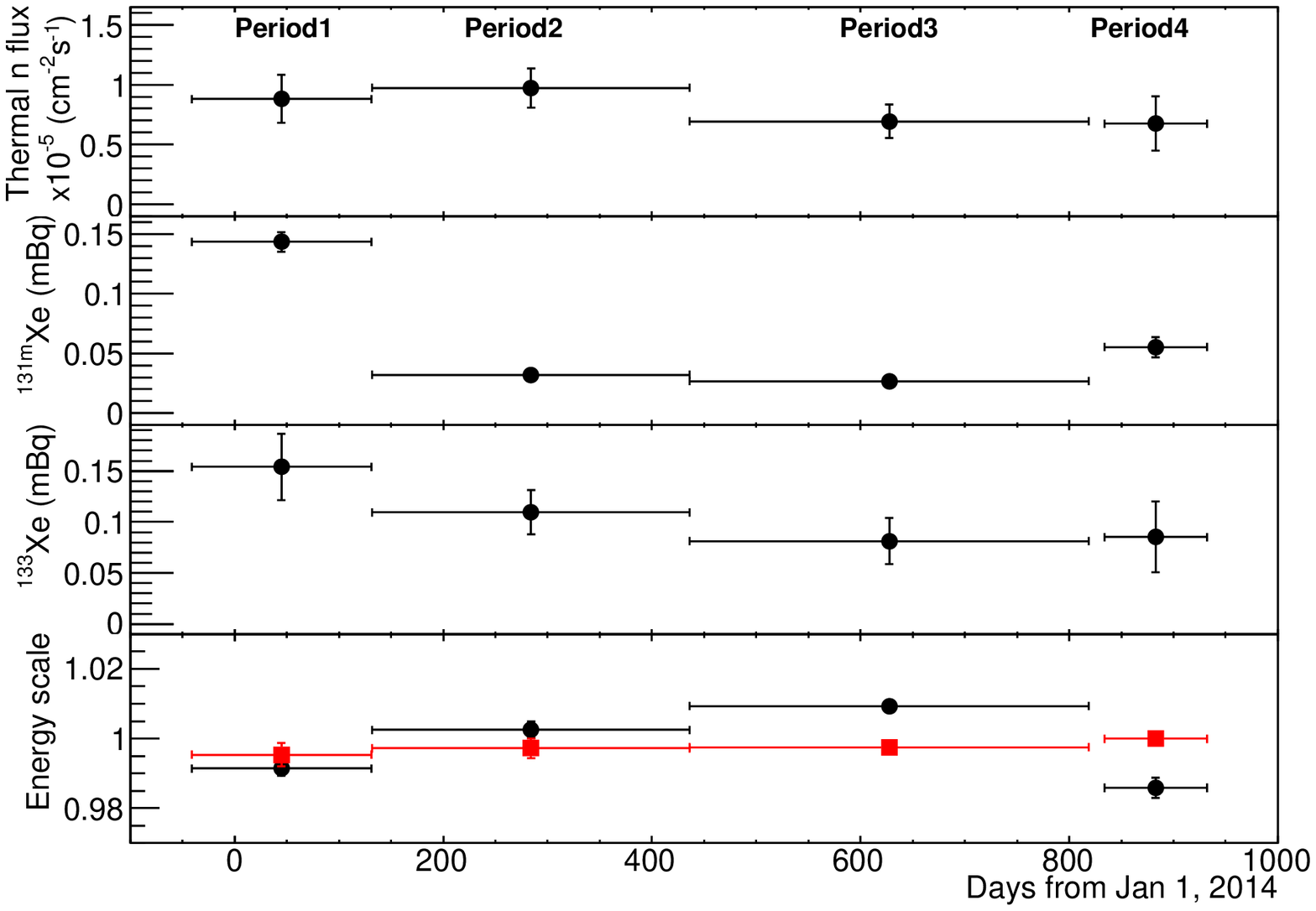}
  \end{center}
  \caption{Time variation of the fitted thermal neutron flux, additional activities of
  $^{131\rm m}$Xe, $^{133}$Xe in the active 832~kg LXe volume, and energy scales
  for the $\beta$-enriched (black circle) and $\beta$-depleted samples (red rectangle).
  The left and right edges of the horizontal error bars represent the start and end of each period,
  respectively.
  }
  \label{fig:fit-result-parameters2.eps}
\end{figure}

Figure~\ref{fig:fit-result-parameters2.eps} shows the variation in time of the fitted thermal neutron flux,
additional activities of $^{131\rm m}$Xe and $^{133}$Xe in the active 832~kg LXe volume, and energy scales.
The fitted thermal neutron flux is stable over the entire data set at $\sim $8$\times 10^{-6}$~/cm$^2$/s.
The larger amounts of $^{131\rm m}$Xe in periods 1 and 4 are explained by
the neutron activation of LXe while storing the LXe outside the water shield
and caused by the $^{252}$Cf calibrations.
Increases in the $^{133}$Xe yield in periods 1 and 4 are not significant
compared with the increases in the $^{131\rm m}$Xe yield.  
The fitted energy scales for the $\beta$-enriched and the $\beta$-depleted samples vary within $\pm$2\%,
which is consistent with the evaluation before fitting. 

\begin{figure}[tbp]
  \begin{center}
    \includegraphics[keepaspectratio=true,width=155mm]{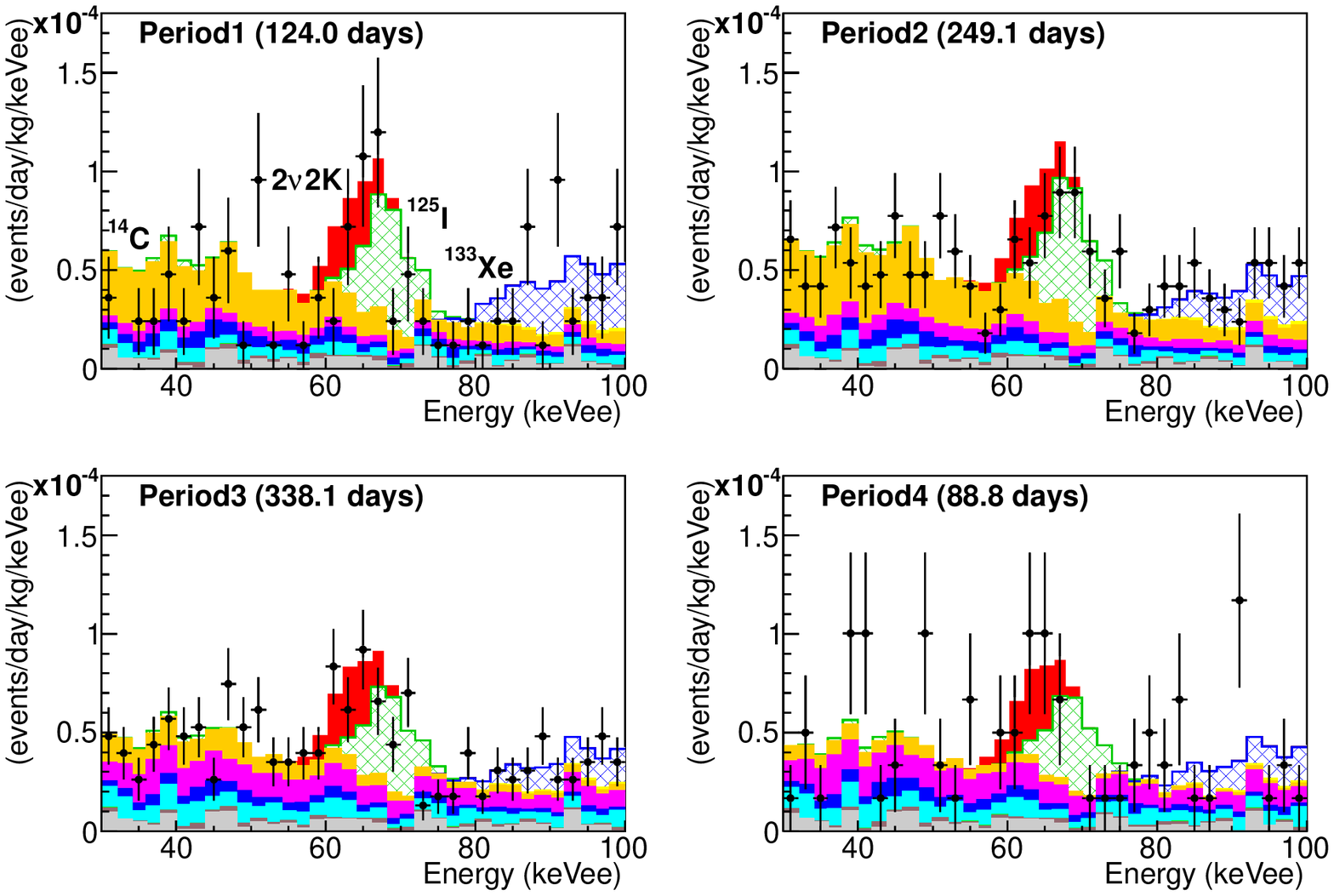}
  \end{center}
  \caption{Closeup figures of energy spectra between 30 and 100~keV$_{\rm ee}$
  for the $\beta$-depleted samples.
  The observed spectra (points) are overlaid with the best-fit
  2$\nu$2K signal and background spectra (colored stacked histograms).
  Colored histograms are the 2$\nu$2K signal (red filled),
  $^{125}$I (green hatched), $^{133}$Xe (blue hatched), $^{14}$C (orange filled), $^{39}$Ar (magenta filled),
  $^{85}$Kr (blue filled), $^{214}$Pb (cyan filled), $^{136}$Xe 2$\nu \beta \beta$ (brown filled),
  and external backgrounds (gray filled).
  }
  \label{fig:fit-result-spectra2.eps}
\end{figure}

Closeup figures of energy spectra between 30 and 100~keV$_{\rm ee}$
for the $\beta$-depleted samples are shown in Fig.~\ref{fig:fit-result-spectra2.eps}.
The peak found at 67.5~keV$_{\rm ee}$ is attributable to the $^{125}$I decay.
The event rate of the $^{125}$I decay is constrained by the thermal neutron flux.

\begin{figure}[tbp]
  \begin{center}
    \includegraphics[keepaspectratio=true,height=85mm]{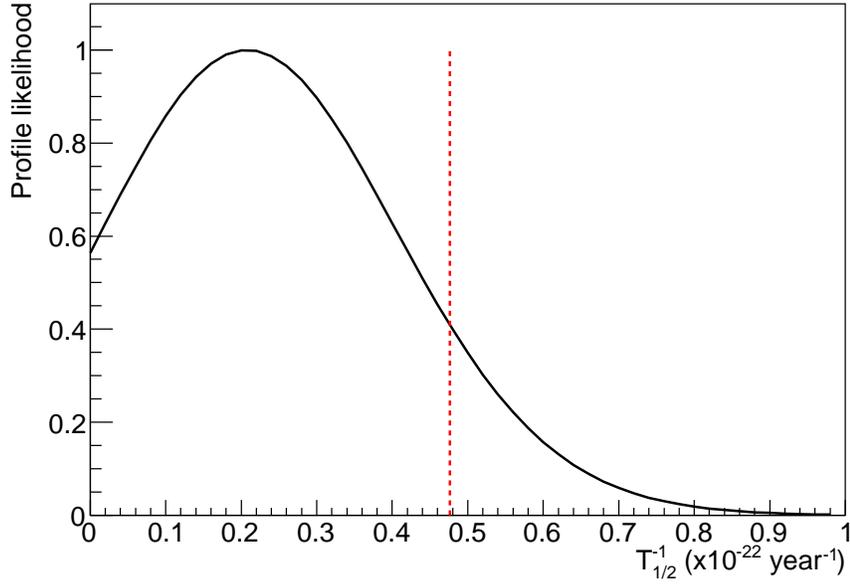}
  \end{center}
  \caption{Normalized profile likelihood $L/L_{\rm max}$ as a function of
  the inverse of the $^{124}$Xe 2$\nu$2K half-life.
  The vertical line indicates the 90\% quantile from which the lower limit
  on the half-life is derived.}
  \label{fig:fit-result-likelihood.eps}
\end{figure}

Figure~\ref{fig:fit-result-likelihood.eps} shows the normalized profile likelihood $L/L_{\rm max}$
as a function of the inverse of the $^{124}$Xe 2$\nu$2K half-life,
where $L_{\rm max}$ is the maximum value of the likelihood.
No significant excess over the expected background is found in the signal region.
We calculate the 90\% CL limit from the relation
\begin{equation}
\frac{\int_{0}^{\xi_{\rm limit}} L(\xi) {\rm d} \xi}
{\int_{0}^{\infty} L(\xi) {\rm d} \xi} = 0.9 \ ,
\end{equation}
where $\xi = 1/T_{1/2}^{2\nu2K}$. This leads to
\begin{equation}
  T_{1/2}^{2\nu 2K} \left (^{124}{\rm Xe} \right ) > \frac{1}{\xi_{\rm limit}} = 2.1 \times 10^{22} \ {\rm years}.
\end{equation}
The fact that we do not observe significant excess above background allows us to give
a constraint on 2$\nu$2K on $^{126}$Xe in the same manner:
\begin{equation}
  T_{1/2}^{2\nu 2K} \left (^{126}{\rm Xe} \right ) > 1.9 \times 10^{22} \ {\rm years}
\end{equation}
at 90\% CL.

\begin{figure}[tbp]
  \begin{center}
    \includegraphics[keepaspectratio=true,height=85mm]{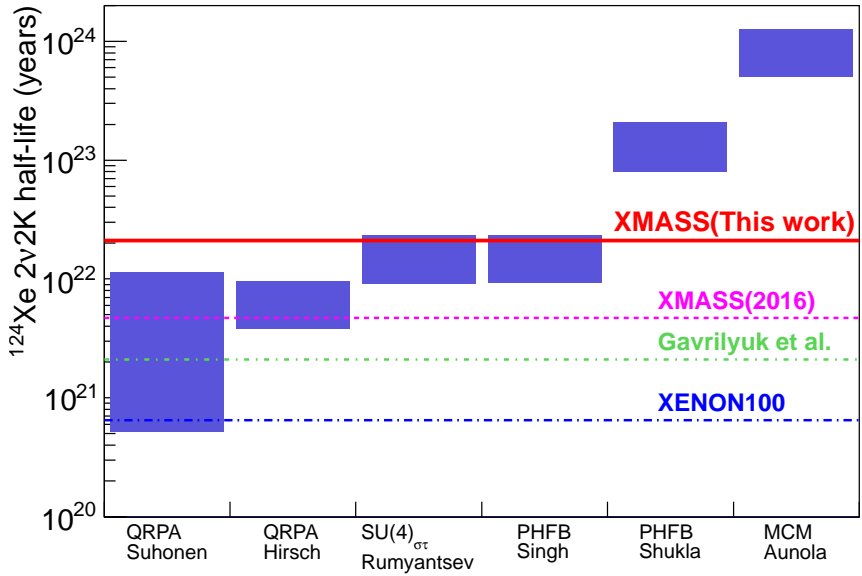}
  \end{center}
  \caption{Comparison of the experimental 90\% CL exclusion limits on
  the $^{124}$Xe 2$\nu$2K half-life overlaid with the theoretical
  calculations~\cite{Hirsch:1994es,Aunola:1996ui,Rumyantsev:1998uy,Shukla:2007ju,Singh:2007jh,Suhonen:2013rca}.
  The lower and upper edges of the theoretical predictions correspond to $g_A=1.26$ and $g_A=1$, respectively.}
  \label{fig:limits-vs-prediction.eps}
\end{figure}

Figure~\ref{fig:limits-vs-prediction.eps} shows a comparison of the experimental 90\% CL
exclusion limits on $^{124}$Xe 2$\nu$2K half-life overlaid with the theoretical
calculations~\cite{Hirsch:1994es,Aunola:1996ui,Rumyantsev:1998uy,Shukla:2007ju,Singh:2007jh,Suhonen:2013rca} for comparison.
The present result gives a lower limit stronger by a factor 4.5 over our previous result,
and gives the most stringent experimental constraint reported to date.
For the theoretical predictions, the reported 2$\nu$ECEC half-lives are converted to 
2$\nu$2K half-lives, divided by the branching ratio for the two electrons being captured
from the $K$-shell, $P_{2K}$=0.767~\cite{Doi:1991xf}.
The lower and upper edges of the bands correspond to $g_A=1.26$ and $g_A=1$, respectively. 

Note that the predicted half-lives will be longer if quenching of $g_A$ is larger.
These experimental results rule out a part of the relevant range of
the reported half-life predictions, and future experiments with multi-ton LXe targets
will have improved sensitivity to further explore this parameter space.

\section{Conclusion}
We have conducted an improved search for 2$\nu$2K on $^{124}$Xe and $^{126}$Xe
using 800.0 days of data from XMASS-I.
For this search, a novel method to discriminate $\gamma$-ray/$X$-ray or 2$\nu$2K signals
from $\beta$-ray backgrounds using LXe scintillation time profiles was developed.
With spectrum fitting in the energy range from 30 to 200~keV$_{\rm ee}$,
no significant 2$\nu$2K signal appeared over the expected background.
Therefore, we set the most stringent lower limits on the half-lives for these processes
at $2.1 \times 10^{22}$ years for $^{124}$Xe
and $1.9 \times 10^{22}$ years for $^{126}$Xe at 90\% CL.

\ack
We gratefully acknowledge the cooperation of the Kamioka Mining and Smelting Company.
This work was supported by the Japanese Ministry of Education, Culture, Sports, Science and Technology,
Grant-in-Aid for Scientific Research (19GS0204, 26104004, and 16H06004),
the joint research program of the Institute for Cosmic Ray Research (ICRR), the University of Tokyo,
and partially by the National Research Foundation of Korea Grant funded by the Korean Government (NRF-2011-220-C00006).

\bibliographystyle{model1-num-names}
\bibliography{<your-bib-database>}

\end{document}